\documentclass[journal=apchd5,manuscript=article,layout=twocolumn
]{achemso}
\usepackage[version=3]{mhchem} 
\usepackage[dvipsnames]{xcolor}
\usepackage{ulem}

\author{Francis Granger}
\affiliation{Univ. Grenoble-Alpes, CEA, Grenoble INP, IRIG, PHELIQS, NPSC, 38000 Grenoble, France.}
\affiliation{Univ. Grenoble-Alpes, CNRS, Inst. NEEL, 38042 Grenoble, France.}

\author{Fabrice Donatini}
\affiliation{Univ. Grenoble-Alpes, CNRS, Inst. NEEL, 38042 Grenoble, France.}

\author{Edith Bellet-Amalric}
\affiliation{Univ. Grenoble-Alpes, CEA, Grenoble INP, IRIG, PHELIQS, NPSC, 38000 Grenoble, France.}

\author{Kuntheak Kheng}
\affiliation{Univ. Grenoble-Alpes, CEA, Grenoble INP, IRIG, PHELIQS, NPSC, 38000 Grenoble, France.}

\author{Gilles Nogues}
\author{David Ferrand}
\author{Joël Cibert}
\email{joel.cibert@neel.cnrs.fr}
\affiliation{Univ. Grenoble-Alpes, CNRS, Inst. NEEL, 38042 Grenoble, France.}

\title{Cathodoluminescence Study of a Quantum Dot in a Nanowire for Single-Photon Emission}

\begin{document}

\begin{tocentry}
\includegraphics{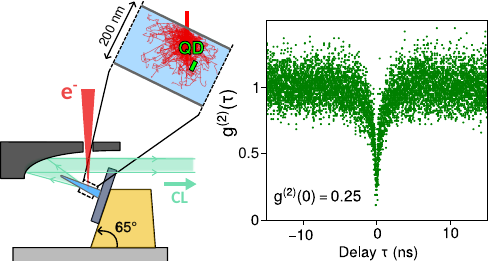} 
\end{tocentry}

\begin{abstract}
Cathodoluminescence in a scanning electron microscope was applied to a semiconductor quantum dot in a nanowire able to emit  single photons. We show that cathodoluminescence can be used not only for imaging and spectroscopy, but also to measure the correlation function and characterize the purity of the single-photon emitter. The electron beam can be manipulated to minimize the collection of parasitic luminescence. At cryogenic temperatures, we observed that the thermal budget, as measured via the phonon sidebands, is close to that of non-resonant micro-photoluminescence. This makes cathodoluminescence an efficient tool in the quest of novel single-photon sources.
\end{abstract}


\section{Introduction}

Sophisticated approaches have been and are currently developed for exploiting quantum optics based on semiconductor quantum dots (QDs). The most popular system is formed by III-V QDs grown by molecular beam epitaxy, for instance InAs QDs grown using the Stransky-Krastanov technique. The neutral or charged exciton emission allows to obtain single photons on-demand with a high purity and brightness, while the biexciton-exciton cascade provides pairs of entangled photons \cite{arakawa2020}. Resonant or quasi-resonant excitation and the insertion in photonic cavities have been particularly refined for QDs emitting in the near infrared \cite{claudon_highly_2010, senellart_high-performance_2017, Tomm2021, Zhang2025} up to the so-called telecom band. \cite{Vajner2024}  \\

In spite of the impressive results achieved so far, there is still a need for further developments  \cite{arakawa2020}. For example, QDs in nanowires are developped in order to facilitate their insertion into a photonic circuit \cite{laferriere_position-controlled_2023-1}. New materials are studied in order to obtain non-cryogenic single-photon emission \cite{Bounouar2012, Deshpande2014, Zhou2018, Zeng2022, Chen2022, Morozov2023, Murtaza2023, granger_brightness_2023} or even the creation of entangled photon pairs at moderately low temperature. Emission at shorter wavelength would allow a reduction of the size and weight of optical components for free-space communication, particularly if satellites are involved. Emission in the blue-green range is required to include underwater operation \cite{LI2019220}, and the UV spectral range to take advantage of a very limited solar illumination background and thus allow for daylight operation. \\

These various requirements imply to develop and test emitters based on QDs of other semiconductor materials or other color centers. Electron-beam excitation (cathodoluminescence, CL) offers an easy and fast way to test these new emitters. In addition, compact electron-beam excitation technique, which was originally developed for lasers, can lead to practical devices \cite{Molva1993, Cuesta2022}.

Previous CL studies of single-photon emission were applied to color centers: neutral NV centers in nanodiamonds \cite{tizei_spatially_2013}, GeV and SiV centers in diamond \cite{bourrellier_bright_2016}, unidentified color centers in hBN \cite{Fiedler2023}. A spectacular feature is the ability to adjust the excitation conditions to switch from single-photon emission, characterised by photon antibunching, to a collective behaviour, which is characterised by photon bunching \cite{Fiedler2023}.

\begin{figure*}[ht]
\centering
    \includegraphics[width=0.9\textwidth]{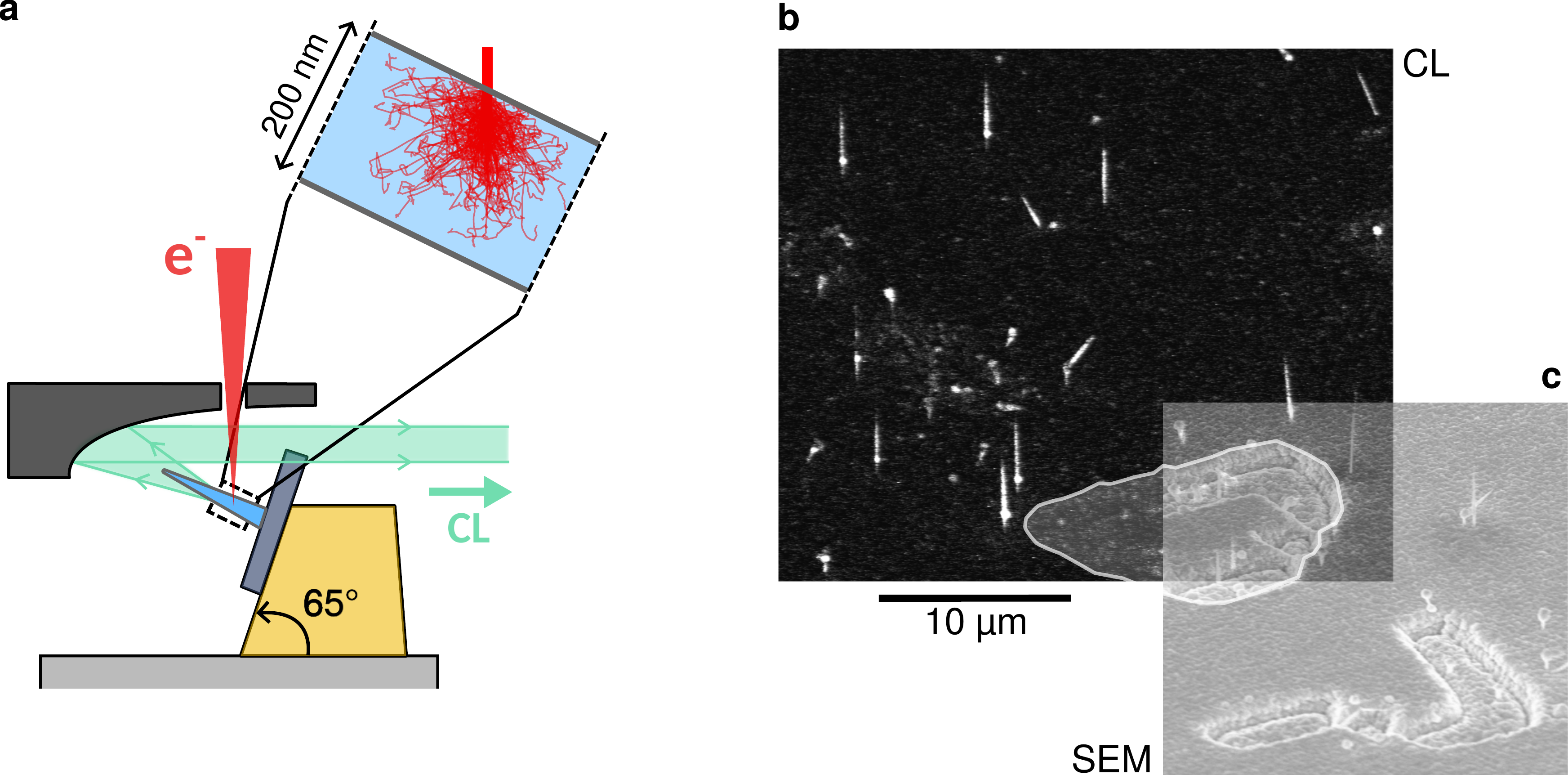}
    \caption{(a) CL setup with inset showing electron trajectories calculated with the CASINO software \cite{drouin_casino_2007} for 5~keV electrons impacting a thick ZnSe slab at 65°. (b) Panchromatic CL image taken with 65° tilt, recorded at room temperature with 5~kV acceleration voltage and 50~pA current. (c) SEM image of the same region, obtained with a Zeiss Ultra+ SEM, partially superimposed on the panchromatic image.}
    \label{figure1}
\end{figure*}

In this paper, we describe the use of a cathodoluminescence setup equipped with a Hanbury-Brown-Twiss (HBT) interferometer to study the single-photon emission characteristics of a single CdSe quantum dot (QD) in a ZnSe nanowire (NW). Section 2 describes the experimental setup, section 3 the cw properties (spectra, diffusion of carriers and its influence on the charged exciton / neutral exciton ratio, and phonon temperature) and section 4 the dynamics (decay and correlations).  \\

\section{Experimental setup}

The ZnSe-CdSe NW-QD sample was grown \cite{gosain_room_2021, Gosain} by molecular beam epitaxy (MBE) on a GaAs (111)B patterned substrate. First, the CdSe QD was embedded near the top of a ZnSe core NW. Typical size is 3~nm in height and 6~nm in diameter \cite{gosain_quantitative_2022}. In a second step of growth, a thick (thickness $\approx200$~nm) and tapered ZnSe shell was formed around the initial NW. The ensemble is typically a 5-6~\textmu m long wire, which acts as an efficient photonic waveguide to increase the collection efficiency along the NW axis \cite{claudon_highly_2010,reimer_bright_2012}. This system has demonstrated bright and single-photon emission at both cryogenic and room-temperature \cite{granger_brightness_2023, gosain_room_2021}. \\

The CL measurements including CL imaging, hyperspectral imaging and time-resolved acquisition are performed in a FEI Inspect F50 SEM equipped with a cryogenic stage. The as-grown QD-NW is placed on a sample holder tilted at 65° so that the light emitted along the NW axis is collected by the parabolic mirror and directed into the spectrometer or the HBT setup \cite{finot_cathodoluminescence_2022}, see Figure~\ref{figure1}(a). The setup used in this work is further detailed in Methods.


\section{Cathodoluminescence from quantum-dots}

\subsection{CL mapping}

Figure~\ref{figure1}(b) presents a wide-field panchromatic CL image obtained at room-temperature, in which the entire luminescence spectrum from the sample is detected by a photomultiplier tube (PMT) using an aluminum mirror instead of a grating within the spectrometer, revealing light-emitting objects. The ZnSe NWs as well as the bent or defective NWs are visible thanks to the luminescence of ZnSe around 460~nm ($\sim2.7$~eV). The bright, round dots at the base of the NWs correspond to the luminescence of QDs emitting in the visible range around 530-580~nm ($2.14-2.34$~eV) at 300~K. The enhanced CL intensity arises from efficient capture of electron–hole pairs by the QD and from the waveguiding effect of the photonic wire optimized for the QD emission \cite{granger_brightness_2023}. Moreover, the collection of light directed along the NW axis is efficiently collected by placing the sample on a tilted support. By contrast, the partially superimposed SEM image (shown in Figure\ref{figure1}(c)) enables us to identify the patterns etched into the GaAs substrate prior to the growth of the NWs, and to come back to a given emitter. \\
		
This image shows that most vertical NWs contain a QD at the base. Therefore, this technique is a simple method to quickly identify promising candidates and make statistical analysis to optimize the growth process.\\

\begin{figure*}[ht]
\centering
    \includegraphics[width=\textwidth]{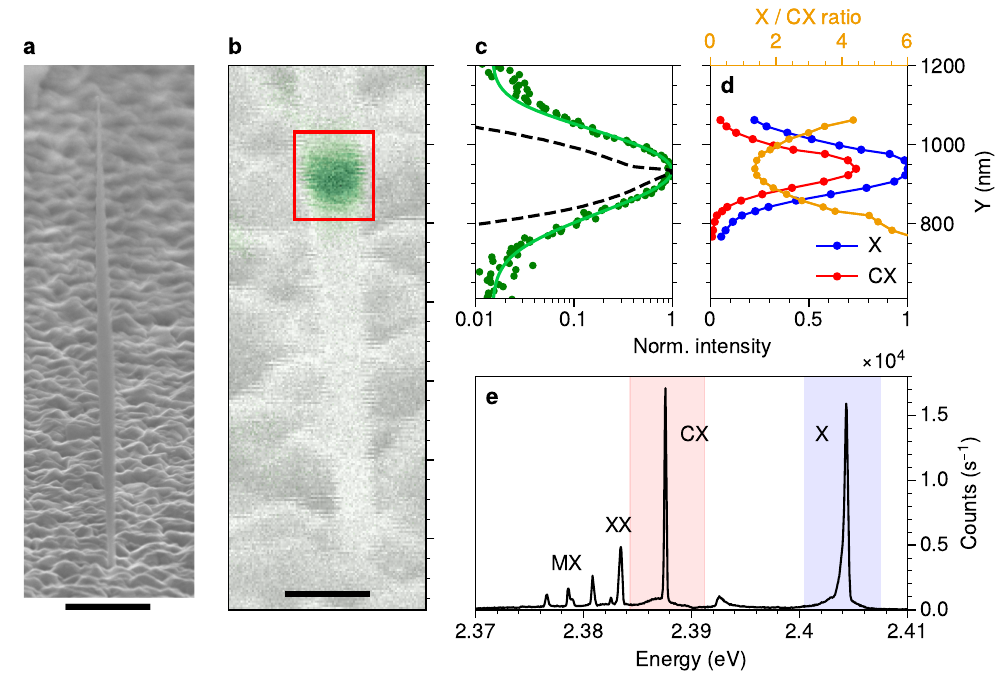}
    \caption{(a) Preliminary SEM image of the NW taken with $80^\circ$ tilt (Zeiss Ultra+). The scale bar shows 1~\textmu m. (b) SEM image of the base of the emitter. The y-axis length is adjusted to account for the $65^\circ$ sample tilt. The corresponding CL map, recorded at 2.39~eV, is superimposed in green. The scale bar shows 200~nm. (c) CL profile (green symbols) obtained from (b) by integrating the CL signal over a 120-nm width across the NW, and fitted by an asymmetric normal-Laplace distribution. The vertical scale is the position along the NW, taking into account the 65° tilt. (d) Exciton (X) and charged exciton (CX) intensities integrated from the spectra shown in (e), and X/CX intensity ratio. (e) Spectrum measured at 5~K from the e-beam scan over the red area indicated in panel (b). We used 1800 grooves/mm, slits 0.05 mm, 5~s exposure time, 5~kV acceleration voltage and 21~pA current.}
    \label{figure2}
\end{figure*}

The SEM image of a selected emitter (to be studied in details below) is displayed in Figure~\ref{figure2}(a). The NW is $5.7$~\textmu m in length and $200$~nm at maximum diameter. The CL signal from this CdSe QD is shown in Figure~\ref{figure2}(b) in green, superimposed on the SEM signal in gray. This signal was obtained under 5~kV acceleration voltage and at cryogenic temperature (5~K) by filtering with a monochromator the CL signal in a spectral window of width  $\sim25$~meV centered around the QD emission energy ($\sim2.39$~eV). The CL image of the QD is superimposed on the SEM image acquired in a separate, sequential scan, allowing us to precisely locate the QD position along the NW - actually the area where the electron beam creates electron-hole pairs which subsequently diffuse into the QD. Similar CL maps were obtained on other NWs on the same sample, including at room-temperature (not shown). The CL intensity profile along the NW axis, obtained by averaging the CL signal over a 120-nm-wide region, is shown in Figure~\ref{figure2}(c) (green symbols). \\

At an acceleration voltage of 5~kV, electrons incident on the NW at a $65^\circ$ angle relative to the NW axis are expected to be slowed down completely within the ZnSe NW. This was confirmed by a Monte-Carlo calculation (CASINO v2.51 \cite{drouin_casino_2007}) performed on a thick ZnSe layer. The calculated trajectories are displayed in the inset of Figure~\ref{figure1}(a): they all stop within 200~nm from the surface. The corresponding distribution of energy deposited by the electrons near the QD, integrated over the NW width and thickness at each position along the NW axis, is plotted as the black dashed line in Figure~\ref{figure2}(c). The calculated distribution is significantly narrower ($\text{FWHM}\approx60$~nm) than the measured CL intensity profile ($\text{FWHM}\approx140$~nm), indicating that carrier diffusion takes place in the QD excitation process.\\

\begin{figure*}[ht]
\centering
    \includegraphics[width=\textwidth]{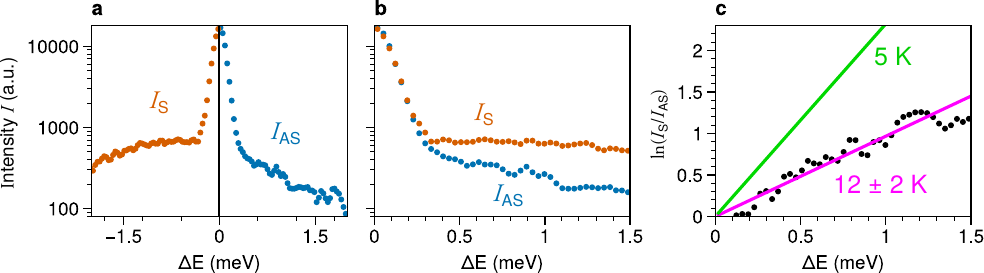}
    \caption{
      (a) Intensity spectrum of a CdSe/ZnSe QD-NW at $T_\mathrm{c ryo}=5\,\mathrm{K}$ under electron-beam excitation (current of 21~pA). Stokes ($I_{\text{S}}$, orange) and anti-Stokes ($I_\text{AS}$, blue) intensities are separated by the ZPL (black line). (b) Stokes and anti-Stokes phonon sidebands. (c) Temperature extraction from the ratio $I_\text{S}/I_\text{AS}$.
      }
    \label{figure3}
\end{figure*}

The CL intensity profile along the NW axis in Figure~\ref{figure2}(c) (green symbols) is well described by an asymmetric normal-Laplace distribution. This shape results from the convolution of an exponential decay (carrier diffusion) with a Gaussian function (electron interaction volume) \cite{artioli_magnetic_2016}. The corresponding fit of the data (green solid line in Figure~\ref{figure2}(c)) gives diffusion lengths of 30~nm below the QD and 40~nm above it. The longer diffusion length observed above the emitter correlates with the larger NW diameter in this region. A similar trend was reported by Artioli \textit{et al.}~\cite{artioli_magnetic_2016} for CdTe/ZnTe QD–NWs, where thicker NW regions exhibited longer carrier diffusion lengths. This is attributed to a smaller impact of surface recombination in wider NW segments.

\subsection{CL spectra}

The CL spectrum displayed in Figure~\ref{figure2}(e) is obtained at a cryostat temperature $T_\text{cryo}=5$~K by scanning the electron beam over a fixed area of nominally $200~\text{nm} \times 220~\text{nm}$ centered on the QD, as illustrated by the red rectangle in Figure~\ref{figure2}(b). Each excitonic line measured at 5~K is composed of a sharp and intense zero-phonon line (ZPL) accompanied by a broad acoustic phonon sideband \cite{besombes_acoustic_2001}. By comparison with other reference emitters from the same sample and under optical excitation \cite{granger_brightness_2023}, the main line around $2.404$~eV is attributed to the neutral exciton (X). The second most intense contribution, located around $2.388$~eV, is attributed to a charged exciton (CX) and the peak around $2.384$~eV to the bi-exciton (XX). The peaks located at low energies are attributed to higher order complexes such as charged bi-exciton or multi-exciton recombinations. The spectrum composition and the intensity of each line agrees with PL measurements on the same sample. The emitter studied here shows a large X-XX separation of $\sim21$~meV, making it a promising candidate for elevated temperature operation. The larger width of the neutral exciton line, as compared to the charged exciton one, is attributed to the presence of (unresolved) fine structure splitting.\\

 We then measured spectra at different excitation spot positions along the NW axis (hyperspectral imaging \cite{bourrellier_bright_2016}). The intensities of the exciton (charged exciton) line, integrated over the blue (red) window in Figure~\ref{figure2}(e), are plotted with respect to the excitation position in Figure~\ref{figure2}(d). Both X and CX lines show maximum intensity when the e-beam spot is centered on the QD. However, the intensity ratio CX/X decreases from $\text{X/CX}=1.3$ at the QD position to $\text{X/XX}>5$ when the excitation spot is displaced by 100~nm. Hence the charged exciton creation involves, at least partially, carriers excited by the electron beam, which diffuse towards the QD with a diffusion length similar to that of the electron-hole pairs.\\

 \subsection{Phonon temperature under e-beam excitation}

The local temperature around the emitter can be extracted from the ratio of the Stokes and anti-Stokes components of the acoustic phonon sideband \cite{granger_calibration-free_2025} which is clearly visible at the base of the X, XX and CX lines in Figure~\ref{figure2}(e). The measure requires no calibration against a temperature standard, and no specific knowledge of the sample details (nature of semiconductor, phonon dispersion, QD size and shape, etc.). The method is analog to temperature measurement through electron energy loss spectroscopy \cite{kikkawa_optical_2022}, but here we measure the temperature of acoustic phonons relevant for the emitter properties. The method is particularly accurate at cryogenic temperatures. If a thermal equilibrium amongst phonons is achieved with a temperature $T_\text{ph}$, the Stokes ($I_\text{S}$) to anti-Stokes ($I_\text{AS}$) intensity ratio is given by:

\begin{equation}
    \frac{I_\text{S}(\Delta E)}{I_\text{AS}(\Delta E)}= \frac{I(-\Delta E)}{I(\Delta E)}=\exp\bigg(\frac{\Delta E}{k_\text{B}T_\text{ph}}\bigg),
\end{equation}

where $\Delta E$ is the energy shift from the ZPL resulting from set of phonons absorbed and emitted, and $k_\text{B}$ the Boltzmann constant.\\

Using this method, as summarized in Figure~\ref{figure3}, we successfully measured the phonon temperature around the CdSe QD under e-beam excitation. The CL spectrum of the CX line measured at $T_\mathrm{c ryo}=5\,\mathrm{K}$ with a current of 21~pA and 5~kV acceleration voltage, is presented in Figure~\ref{figure3}(a). At this cryogenic temperature, the phonon population is small. As a result, the ZPL is more intense and narrower than the phonon sideband, and one notices a strong asymmetry between the Stokes (net phonon emission, lower photon energy) and anti-Stokes (net phonon absorption, higher photon energy) sideband components.\\

The intensities $I_\text{S}$ and $I_\text{AS}$ are plotted as a function of the shift with respect to the ZPL position in Figure~\ref{figure3}(b). The logarithm of the Stokes to anti-Stokes ratio ($I_\text{S}/I_\text{AS}$) is
displayed in Figure~\ref{figure3}(c), covering energy shifts $\Delta E$ up to $1.5$~meV. The linear fit of the data provides a phonon temperature of $T_\text{ph}=12\pm2$~K, significantly higher than the cryostat temperature $T_\text{cryo}=5$~K. 

This value is similar to the temperature measured under non-resonant photo-excitation of emitters from the same sample \cite{granger_calibration-free_2025}, with a temperature increase attributed to the laser heating effect.

\section{Time-resolved cathodoluminescence}

\begin{figure*}[ht]
    \centering
    \includegraphics[width=\textwidth]{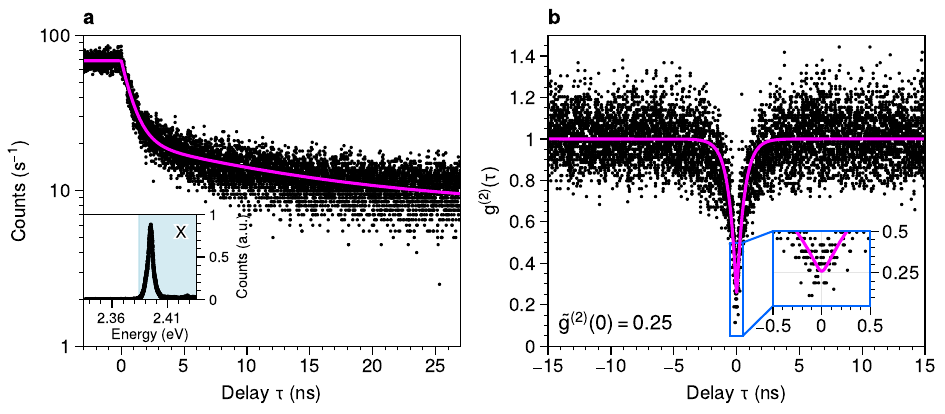}
    \caption{(a) Decay time of the neutral exciton acquired at 5 K using 5 kV acceleration voltage and a current of  960 pA with 1 MHz excitation rate. After subtraction of a constant baseline, the solid magenta line shows bi-exponential decay curve with fast decay time $\tau_X=0.9$~ns and a longer one of $14$~ns. The inset shows the exciton emission spectrum recorded with a tunable filter and a 7~mm slit width, under the same beam current and acceleration voltage conditions. (b) Exciton autocorrelation (acquisition time 600~s, 5 kV acceleration voltage, 960 pA beam current) obtained from an e-beam scan over a fixed area similar to the red rectangle in Figure~\ref{figure2}(b). The magenta curve shows a fit based on the convolution of a Laplace distribution with the HBT instrumental response, with the adjustable parameters $T_\text{CL} = 0.6$~ns and $B/S\approx0.14$.}
    \label{figure4}
\end{figure*}

The dynamics of the emitter can also be studied using the e-beam excitation. In this section, we used a larger spot-size than in the previous one; thus increasing the e-beam current from 21~pA to  960~pA. The other conditions are maintained, acceleration voltage of 5~kV, cryostat temperature 5~K. The inset of Figure~\ref{figure4}(a) shows the neutral exciton line, recorded with the electron beam scanned over a fixed region centered on the QD (see the red rectangle shown in Figure~\ref{figure2}(b)). To record the time-dependent signals, we used a tunable bandpass filter that only transmits the exciton emission line, as schematized by the blue area in the inset of Figure~\ref{figure4}(a) .

\subsubsection{Decay time}

The decay of the exciton intensity is presented in Fig~\ref{figure4}(a). Pulsed excitation is achieved by rapidly displacing the electron beam in the direction perpendicular to the NW axis \cite{donatini2018exciton}, using a beam blanker at a frequency of 1~MHz. The duty cycle was 50\% and the switching time was smaller than 70~ps.  \\

The signal decay is well fitted by a bi-exponential function. The exciton features a dominant fast decay component with a characteristic time of $\tau_X=0.9$~ns, accompanied by a weaker component, fitted at $14$~ns but which may expand to longer delays. These two time constants are of the same order of magnitude as those (0.6~ns and 30~ns) measured on another NW-QD by time-resolved PL \cite{gosain_quantitative_2022}.

\subsubsection{Autocorrelation}

Figure~\ref{figure4}(b) shows the normalized second-order autocorrelation function $g^{(2)}(\tau)$ measured under continuous excitation on the neutral exciton line. This measurement was performed simultaneously with the acquisition of the photoluminescence spectrum presented in the inset of Figure~\ref{figure4}(a). A clear antibunching behavior is observed, with a zero-delay $\tilde{g}^{(2)}(0)$ well below the threshold of 0.5, thus confirming single-photon emission.\\  

To our knowledge, this is the first observation of photon antibunching from a single QD measured via cathodoluminescence. Previous CL measurements of the correlation function were obtained on color centers, 
with $g^{(2)}(0)$ values around 0.5 for NV centers in nanodiamond ~\cite{tizei_spatially_2013} or point defects in \textit{h}-BN \cite{bourrellier_bright_2016}. A lower value, $g^{(2)}(0)=0.06$, was reported for Ge-V and Si-V centers in diamond \cite{Fiedler2023}, using a luminescence filtered by a spectrometer and a detailed fit of the $g^{(2)}$ function. \\

Remarkably, our result was obtained over the full excitonic emission line using the broad bandpass filter. Figure~\ref{figure4}(a) suggests that a long-time contribution is present, in addition to the fast excitonic decay at $\tau_X=0.9$~ns. Further studies are needed to decide whether this signal is due to excitation of traps, either through the luminescence of these traps, or by the re-excitation of the QD from these traps, since these two mechanisms will have a different impact on the correlation function.\\

Such a correlation function is generally described by a Laplace distribution~\cite{brouri_photon_2000, gosain_quantitative_2022, geraci_family_2020,tizei_spatially_2013} with a characteristic time $T_{\text{CL}}$. This theoretical outcome results from a simple model of the emitter as a two-level system restricted to the neutral single exciton and bi-exciton\cite{moreau_quantum_2001}. In this model, the characteristic CL photon production rate $1/T_\text{CL}$ is the sum of the excitation rate $p$ and the exciton decay rate $1/\tau_\text{X}$: 

\begin{equation}\label{eq2}
	\frac{1}{T_{\text{CL}}} = p + \frac{1}{\tau_X}.
\end{equation}

The measured autocorrelation function $\tilde{g}^{(2)}$ results from the convolution of this Laplace distribution with the instrumental response of the HBT setup, which is modeled by a Gaussian function with standard deviation $\sigma$. The response time of the two arms of the HBT setup was fixed at $\sigma = 30~\text{ps}$, corresponding to a FWHM response time of each detector of $50~\text{ps}$. The fit in Figure~\ref{figure4}(b) was obtained with $T_{\text{CL}} = 0.6~\text{ns}$, and $\tilde{g}^{(2)}(0)=0.25$.\\



From the spectrum shown in the inset of Figure~\ref{figure4}(a), the contribution of bi-exciton emission to $g^{(2)}(0)$ is expected to be negligible. The expected reduction of the zero-delay value due exclusively to timing jitter is $\sqrt{\frac{2}{\pi}}\frac{\sigma}{T_{\text{CL}}} \approx 0.04$. The remaining deviation is thus attributed to the presence of uncorrelated background, which leads to a nonzero value of $\tilde{g}^{(2)}(0)$. To account for the signal-to-background ratio, a prefactor $\frac{S}{S+B}$ is introduced \cite{brouri_photon_2000}, where $S$ and $B$ denote the signal and background contributions, respectively. The fit in Figure \ref{figure4}(b), assuming an ideal single-photon emitter, yields an uncorrelated background signal $B/S \approx 0.14$. This level is comparable to values reported for other samples under non-resonant optical excitation~\cite{gosain_quantitative_2022}, and is remarkably low given the broad ($\sim 0.1$~eV) spectral detection window used in the experiment.\\ 

The characteristic time $T_{\text{CL}} = 0.6~\text{ns}$ is shorter than the exciton decay time $\tau_X = 0.9~\text{ns}$ obtained from Figure~\ref{figure4}(a). Eq.~\ref{eq2}  leads to $p\tau_X=0.5$, meaning that the system is approximately half-way from saturation, in spite of the large spot size used in our time-resolved study. We note however that the excitation by an electron beam should be addressed specifically, as each incident electron creates several hundred electron-hole pairs. This aspect is beyond the scope of the present paper and only briefly addressed in the discussion.\\


\section{Discussion}


Exploring new systems and new materials for single-photon emission requires using non-resonant excitation at least in the early stage of development. An important issue is the thermal budget. In the case of laser excitation of a QD in a nanowire, nearly all absorbed photons above the bandgap contribute directly to electron–hole pair creation in the nanowire material. Only electron-hole pairs created close enough to the QD contribute to the single-photon emission, while a good part of the rest contribute to heating the environment through non-radiative processes \cite{granger_calibration-free_2025}. This is particularly crucial at low temperature, as the thermal conductivity varying as $T^3$, becomes much smaller. Heating also affects non-resonant excitation of Stranski-Krastanov QDs since electron-hole pairs are created in a micrometer-sized volume in the barrier material. Of course, heating is drastically reduced in the case of resonant or quasi-resonant excitation  \cite{granger_calibration-free_2025} but such methods cannot be used in the first stage of a study.

In the case of e-beam excitation, only a fraction of the incident electrons give rise to electron-hole pairs capable of exciting the QD. This is because the primary high-energy electrons mainly lose energy through inelastic scattering processes that produce secondary electrons, Auger electrons, X-rays, electron-hole pairs and phonons. In bulk materials or thick layers, the rule of thumb is that typically 70\% of the deposited energy is dissipated as heat \cite{Colak1985}. This disadvantage with respect to the laser beam, is compensated here by the small size of the excitation volume, so that the energy is deposited in the close vicinity of the QD, leading to an efficient carrier capture. Dedicated studies of the dependence on acceleration voltage and beam current would be needed to identify the thermal channels, on one hand, and to understand the mechanisms of excitation of a QD by an electron beam, on the other hand.\\

Another interesting aspect is the possibility to adjust the excitation volume by choosing the acceleration voltage. At 5 kV, the acceleration voltage used in the present study as in Ref.\cite{Fiedler2023}~, the incident electrons are stopped in the nanowire if the beam is centered on it. That implies that several hundreds of electron-hole pairs are created by each incident electron: this specificity is currently exploited to create collective excitations in an ensemble of localized centers \cite{Fiedler2023}. This should be taken into account in the excitation process of a single-photon emitter: at a current of 20~pA (\textit{i.e}., 1.25$\times{10^8}$ incident electrons per second), this big electron-hole packet is created in average every 8~ns, ten times longer than the exciton lifetime. Other studies have used fast electrons, for instance 60-100 keV in nanocristallites \cite{tizei_spatially_2013}) with the idea that only a small part of the incident energy is deposited in the nano-object and at most one electron-hole pair around the bandgap is created at each incident electron.

Using a small acceleration voltage also limits the contribution of other emitters nearby. This is quite efficient when the beam is static on the nanowire. The case of a scanned e-beam is not as simple. 

An interesting aspect is that the cathodoluminescence measurement of the purity could even replace the preliminary micro-photoluminescence characterization and allow for a single-step fabrication of the single-photon emitting device using e-beam lithography \cite{Donatini2010}.\\


Unlike optical techniques, the electron microscope gives access to a high-resolution image of the structure. In addition, the sub-wavelength resolution of the e-beam excitation allows to measure variations in $g^{(2)}$ as we scan the e-beam around the emitter. This advantage over optical excitation is even more remarkable when emitters are separated by only few nanometers, as demonstrated on $\text{NV}^0$ centers by Tizei \textit{et al.}~\cite{tizei_spatially_2013}. The sample presented in our work contains QD-NWs separated by few micrometers and each NW contains a single QD. Therefore multi-photon emission involving more than one QD is very unlikely via e-beam excitation. 

As shown in Figure~\ref{figure2}(d), we found that the charge state of the QD varies as the excitation spot moves away from the QD. This approach enables us to probe the quantum properties of the same emitter in different charge configurations.

\section{Conclusion}

Cathodoluminescence is already well-known and widely used as the primary tool for characterizing luminescence from localized emitters, such as QDs. We show that it presents a thermal budget equivalent to non-resonant photoexcitation, and offers excellent prospects for simultaneously studying correlation properties and characterizing the purity of a QD single-photon emitter. This makes cathodoluminescence a valuable tool for developing new single-photon sources,  as those required by specific applications beyond the near-infrared fiber-based configuration of quantum communication. 

\section{Methods}
\label{sec:methods}

\begin{figure*}[h]
    \centering
    \includegraphics[width=14cm]{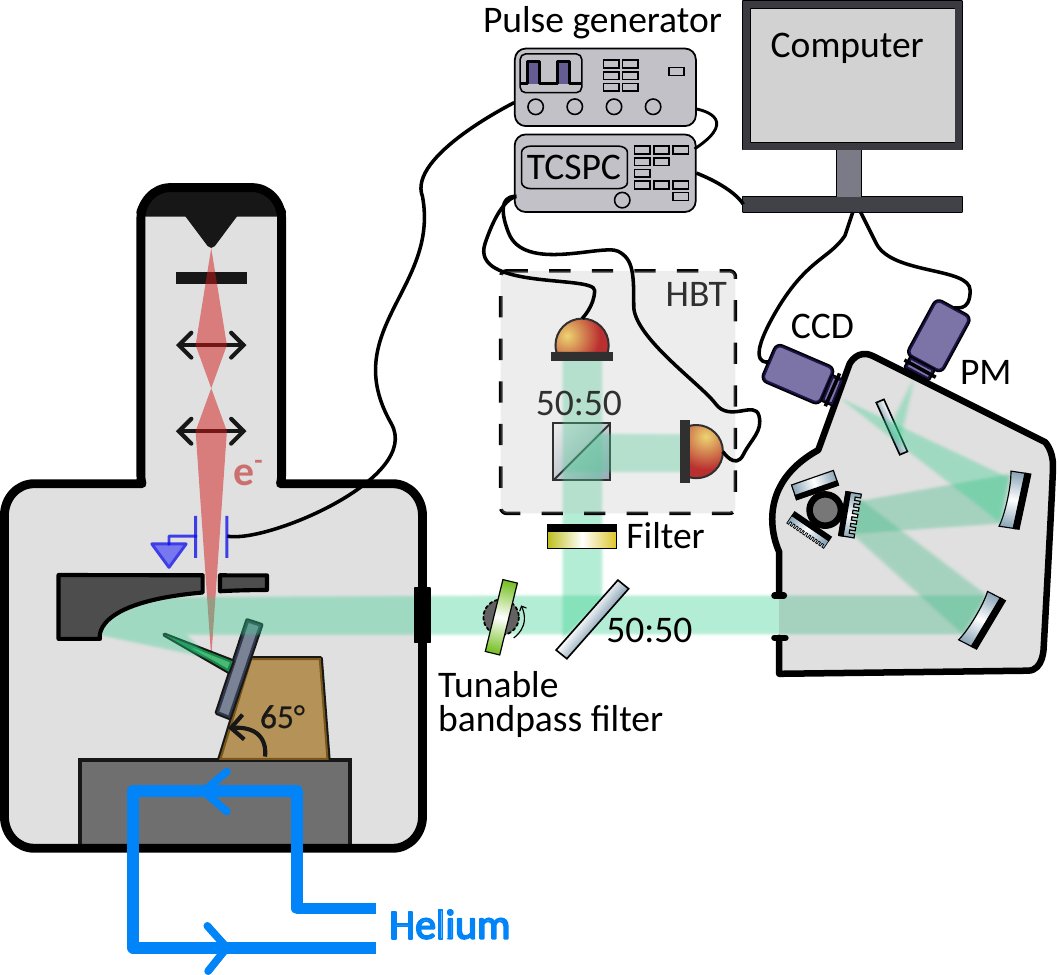}
	\caption{Schematic of the SEM setup used for CL and TRCL measurements.}
\label{fig:CL_setup}
\end{figure*}

\subsection{Cathodoluminescence setup}

A schematic of the CL measurement setup used in this work is shown in Figure~\ref{fig:CL_setup}. The CL setup at Institut Néel consists of a custom FEI Inspect F50 SEM, equipped with a cryogenic stage allowing the sample to be theoretically cooled down to 5~K thanks to the circulation of liquid helium.\\

The studied QD-NW is placed on a tilted sample holder (65°) and the light emitted along the NW axis is collected by a parabolic mirror and directed into a spectrometer (Horiba iHR550). The spectrometer is equiped with two gratings (600 tr/mm and 1800 tr/mm) and a mirror. The signal can be analysed with a CCD or a photomultiplier tube, allowing CL spectra and images to be acquired with SEM resolution. \\
  
The setup allows to perform time-resolved cathodoluminescence (TRCL). The generation of electron pulses is performed using the beam blanking technique \cite{donatini2018exciton} and consists in inserting one pair of electrostatic deflector plate into the microscope column. By applying a bias, one can deflect the e-beam. Therefore,  pulse frequency and its duty cycle can be easily tuned using a pulse generator. However, the beam blanker introduces astigmatism and the temporal and spatial resolution is limited by the displacement of the e-beam on the sample during the switching.\\	

The setup is equipped with an HBT setup which was developed by \cite{finot_cathodoluminescence_2022, Finot2022}. Autocorrelation can be performed with two fast PMTs, with a time resolution better than 50 ps (PMA Hybrid 06 for PicoQuant), adapted for UV emission \cite{finot_cathodoluminescence_2022}. The fast pulse generator (PG-1072 rev. B from Active Technologies) whichs drives the beam blanker is also used as a trigger for the TCSPC photon counting device (PicoHarp 300 from PicoQuant). \\
				
The analysis of a single emission line can be performed thanks to the tunable bandpass filter (TBP01-561 from Semrock). The HBT setup allows one to characterize the single-photon character of the source and its lifetime. Finally, the optical measurements using the spectrometer can be performed during HBT acquisition thanks to the the 50:50 plate placed in front of the two setups. \\

\begin{acknowledgement}

We thank Gwénolé Jacopin for giving us the opportunity to use the HBT setup.\\

We acknowledge funding from LANEF in Grenoble, ANR-10-LABX-51-01, and CEA-PE Bottom-Up QPhotonics.

\end{acknowledgement}




%

\bibliography{references}

@phdthesis{artioli_magnetic_2016,
	type = {Ph.{D}. thesis},
	title = {Magnetic polaron in ({Cd},{Mn}){Te} quantum dot inserted in {ZnTe} nanowire},
	url = {https://theses.hal.science/tel-01424256},
	urldate = {2025-05-04},
	school = {Université Grenoble Alpes},
	author = {Artioli, Alberto},
	month = jun,
	year = {2016},
}

@phdthesis{finot_cathodoluminescence_2022,
	type = {Ph.{D}. thesis},
	title = {Cathodoluminescence lifetime spectroscopy for efficient {III}-nitride {LEDs}},
	url = {https://theses.hal.science/tel-04002669},
	urldate = {2025-05-08},
	school = {Université Grenoble Alpes},
	author = {Finot, Sylvain},
	month = nov,
	year = {2022},
}

@article{ Finot2022,
Author = {Finot, Sylvain and Le Maoult, Corentin and Gheeraert, Etienne and
   Vaufrey, David and Jacopin, Gwenole},
Title = {Surface Recombinations in III-Nitride Micro-LEDs Probed by
   Photon-Correlation Cathodoluminescence},
Journal = {ACS Photonics},
Year = {2022},
Volume = {9},
Number = {1},
Pages = {173-178},
Month = {JAN 19},
DOI = {10.1021/acsphotonics.1c01339},
EarlyAccessDate = {DEC 2021},
ISSN = {2330-4022},
ResearcherID-Numbers = {JACOPIN, Gwenole/M-7360-2015
   JACOPIN, Gwénolé/M-7360-2015
   Finot, Sylvain/ABE-1173-2021
   Gheeraert, Etienne/E-8071-2015
   },
ORCID-Numbers = {JACOPIN, Gwenole/0000-0003-0049-7195
   Gheeraert, Etienne/0000-0002-9952-5805
   Finot, Sylvain/0000-0003-1455-7160},
Unique-ID = {WOS:000729741600001},
}

@article{granger_brightness_2023,
	title = {Brightness and purity of a room-temperature single-photon source in the blue–green range},
	volume = {48},
	copyright = {© 2023 Optica Publishing Group},
	issn = {1539-4794},
	url = {https://opg.optica.org/ol/abstract.cfm?uri=ol-48-15-3833},
	doi = {10.1364/OL.492039},
	number = {15},
	urldate = {2023-07-18},
	journal = {Optics Letters},
	author = {Granger, Francis and Gosain, Saransh Raj and Nogues, Gilles and Bellet-Amalric, Edith and Cibert, Joël and Ferrand, David and Kheng, Kuntheak},
	month = aug,
	year = {2023},
	note = {Publisher: Optica Publishing Group},
	pages = {3833--3836},
}

@article{Zeng2022,
author = {Helen ZhiJie Zeng and Minh Anh Phan Ngyuen and Xiaoyu Ai and Adam Bennet and Alexander S. Solntsev and Arne Laucht and Ali Al-Juboori and Milos Toth and Richard P. Mildren and Robert Malaney and Igor Aharonovich},
journal = {Opt. Lett.},
number = {9},
pages = {2161--2161},
publisher = {Optica Publishing Group},
title = {Integrated room temperature single-photon source for quantum key distribution},
volume = {47},
month = {May},
year = {2022},
url = {https://opg.optica.org/ol/abstract.cfm?URI=ol-47-9-2161},
doi = {10.1364/OL.460614},
}

@article{Murtaza2023,
author = {Ghulam Murtaza and Maja Colautti and Michael Hilke and Pietro Lombardi and Francesco Saverio Cataliotti and Alessandro Zavatta and Davide Bacco and Costanza Toninelli},
journal = {Opt. Express},
number = {6},
pages = {9437--9447},
publisher = {Optica Publishing Group},
title = {Efficient room-temperature molecular single-photon sources for quantum key distribution},
volume = {31},
month = {Mar},
year = {2023},
url = {https://opg.optica.org/oe/abstract.cfm?URI=oe-31-6-9437},
doi = {10.1364/OE.476440},
}

@article{Zhang2025,
  title = {Experimental Single-Photon Quantum Key Distribution Surpassing the Fundamental Weak Coherent-State Rate Limit},
  author = {Zhang, Yang and Ding, Xing and Li, Yang and Zhang, Likang and Guo, Yong-Peng and Wang, Gao-Qiang and Ning, Zhen and Xu, Mo-Chi and Liu, Run-Ze and Zhao, Jun-Yi and Zou, Geng-Yan and Wang, Hui and Cao, Yuan and He, Yu-Ming and Peng, Cheng-Zhi and Huo, Yong-Heng and Liao, Sheng-Kai and Lu, Chao-Yang and Xu, Feihu and Pan, Jian-Wei},
  journal = {Phys. Rev. Lett.},
  volume = {134},
  issue = {21},
  pages = {210801},
  numpages = {6},
  year = {2025},
  month = {May},
  publisher = {American Physical Society},
  doi = {10.1103/PhysRevLett.134.210801},
  url = {https://link.aps.org/doi/10.1103/PhysRevLett.134.210801}
}

@article{Zhou2018,
author = {Yu Zhou  and Ziyu Wang  and Abdullah Rasmita  and Sejeong Kim  and Amanuel Berhane  and Zoltán Bodrog  and Giorgio Adamo  and Adam Gali  and Igor Aharonovich  and Wei-bo Gao },
title = {Room temperature solid-state quantum emitters in the telecom range},
journal = {Science Advances},
volume = {4},
number = {3},
pages = {eaar3580},
year = {2018},
doi = {10.1126/sciadv.aar3580},
URL = {https://www.science.org/doi/abs/10.1126/sciadv.aar3580},
eprint = {https://www.science.org/doi/pdf/10.1126/sciadv.aar3580}}

@article{Deshpande2014,
    author = {Deshpande, Saniya and Frost, Thomas and Hazari, Arnab and Bhattacharya, Pallab},
    title = {Electrically pumped single-photon emission at room temperature from a single InGaN/GaN quantum dot},
    journal = {Applied Physics Letters},
    volume = {105},
    number = {},
    pages = {141109},
    year = {2014},
    month = oct,
    doi = {10.1063/1.4897640},
    url = {https://doi.org/10.1063/1.4897640},
}

@article{Chen2022,
author = {Chen, Ling and Sheng, Bowen and Sheng, Shanshan and Wang, Ping and Sun, Xiaoxiao and Li, Duo and Wang, Tao and Tao, Renchun and Liu, Shangfeng and Chen, Zhaoying and Ge, Weikun and Shen, Bo and Wang, Xinqiang},
title = {Room Temperature Triggered Single Photon Emission from Self-Assembled GaN/AlN Quantum Dot in Nanowire},
journal = {Advanced Functional Materials},
volume = {32},
number = {47},
pages = {2208340},
doi = {https://doi.org/10.1002/adfm.202208340},
url = {https://advanced.onlinelibrary.wiley.com/doi/abs/10.1002/adfm.202208340},
year = {2022}
}

@article{Bounouar2012,
author = {Bounouar, S. and Elouneg-Jamroz, M. and Hertog, M. den and Morchutt, C. and Bellet-Amalric, E. and Andr{\'e}, R. and Bougerol, C. and Genuist, Y. and Poizat, J.-Ph. and Tatarenko, S. and Kheng, K.},
title = {Ultrafast Room Temperature Single-Photon Source from Nanowire-Quantum Dots},
journal = {Nano Letters},
volume = {12},
number = {6},
pages = {2977-2981},
year = {2012},
doi = {10.1021/nl300733f},
}

@Article{Morozov2023,
author ="Morozov, Sergii and Vezzoli, Stefano and Myslovska, Alina and Di Giacomo, Alessio and Mortensen, N. Asger and Moreels, Iwan and Sapienza, Riccardo",
title ="Purifying single photon emission from giant shell CdSe/CdS quantum dots at room temperature",
journal ="Nanoscale",
year ="2023",
volume ="15",
issue ="4",
pages ="1645-1651",
publisher ="The Royal Society of Chemistry",
doi ="10.1039/D2NR04744F",
}

@article{Tomm2021,
	title = {A bright and fast source of coherent single photons},
	volume = {16},
	url = {},
	doi = {},
	number = {},
	urldate = {},
	journal = {Nature Nanotechnology},
	author = {Tomm, Natasha and Javadi, Alisa and Antoniadis, Nadia Olympia and Najer, Daniel and Löbl, Matthias Christian and Korsch, Alexander Rolf and Schott, Rüdiger and Valentin, Sascha René and Wieck, Andreas Dirk and Ludwig, Arne and Warburton, Richard John},
	month = apr,
	year = {2021},
	note = {},
	pages = {399},
}

@article{Vajner2024,
	title = {On-Demand Generation of Indistinguishable Photons in the Telecom C-Band Using Quantum Dot Devices},
	volume = {11},
	url = {},
	doi = {},
	number = {},
	urldate = {},
	journal = {ACS Photonics},
		author = {Vajner, D},
	month = jan,
	year = {2024},
	note = {},
	pages = {319},
}

@article{claudon_highly_2010,
	title = {A highly efficient single-photon source based on a quantum dot in a photonic nanowire},
	volume = {4},
	copyright = {2010 Springer Nature Limited},
	issn = {1749-4893},
	url = {https://www.nature.com/articles/nphoton.2009.287x},
	doi = {10.1038/nphoton.2009.287x},
	number = {3},
	urldate = {2025-02-26},
	journal = {Nature Photonics},
	author = {Claudon, Julien and Bleuse, Joël and Malik, Nitin Singh and Bazin, Maela and Jaffrennou, Périne and Gregersen, Niels and Sauvan, Christophe and Lalanne, Philippe and Gérard, Jean-Michel},
	month = mar,
	year = {2010},
	note = {Publisher: Nature Publishing Group},
	pages = {174--177},
}

@article{reimer_bright_2012,
	title = {Bright single-photon sources in bottom-up tailored nanowires},
	volume = {3},
	copyright = {2012 The Author(s)},
	issn = {2041-1723},
	url = {https://www.nature.com/articles/ncomms1746},
	doi = {10.1038/ncomms1746},
	number = {1},
	urldate = {2023-03-29},
	journal = {Nature Communications},
	author = {Reimer, Michael E. and Bulgarini, Gabriele and Akopian, Nika and Hocevar, Moïra and Bavinck, Maaike Bouwes and Verheijen, Marcel A. and Bakkers, Erik P. A. M. and Kouwenhoven, Leo P. and Zwiller, Val},
	month = mar,
	year = {2012},
	pages = {737},
}

@article{Fiedler2023,
	title = {Sub-to-super-Poissonian photon statistics in cathodoluminescence of color center ensembles in isolated diamond crystals},
	volume = {12},
	copyright = {},
	issn = {},
	url = {},
	doi = {},
	number = {},
	urldate = {},
	journal = {Nanophotonics},
	author = {Fiedler, Saskia and Morozov, Sergei and Komisar, Danylo  and Ekimov, Evgeny A. and Kulikova, Liudmila F.  and Davydov, Valery A. and Agafonov, Viatcheslav N. and Kumar, Shailesh  and Wolff, Christian and Bozhevolnyi, Sergey I. and Mortensen, N. Asger},
	month = mar,
	year = {2023},
	pages = {2231},
}

@article{Colak1985,
	title = {Electron beam pumped II-VI lasers	},
	volume = {72},
	copyright = {},
	issn = {},
	url = {},
	doi = {},
	number = {},
	urldate = {},
	journal = {J. Cryst. Growth},
	author = {Colak, S. and Fitzpatrick, B. J. and Bhargava, R. N. },
	month = jul,
	year = {1985},
	pages = {504},
}

@article{Cuesta2022,
doi = {10.1088/1361-6463/ac6237},
url = {https://doi.org/10.1088/1361-6463/ac6237},
year = {2022},
month = {apr},
publisher = {IOP Publishing},
volume = {55},
number = {27},
pages = {273003},
author = {Cuesta, Sergi and Harikumar, Anjali and Monroy, Eva},
title = {Electron beam pumped light emitting devices},
journal = {Journal of Physics D: Applied Physics}
}

@article{Molva1993,
    author = {Molva, E. and Accomo, R. and Labrunie, G. and Cibert, J. and Bodin, C. and Dang, Le Si and Feuillet, G.},
    title = {Microgun‐pumped semiconductor laser},
    journal = {Applied Physics Letters},
    volume = {62},
    number = {8},
    pages = {796-798},
    year = {1993},
    month = feb,
    issn = {0003-6951},
    doi = {10.1063/1.108581},
    url = {https://doi.org/10.1063/1.108581},
    
}

@article{drouin_casino_2007,
	title = {{CASINO} {V2}.42: a fast and easy-to-use modeling tool for scanning electron microscopy and microanalysis users},
	volume = {29},
	issn = {0161-0457},
	shorttitle = {{CASINO} {V2}.42},
	doi = {10.1002/sca.20000},
	number = {3},
	journal = {Scanning},
	author = {Drouin, Dominique and Couture, Alexandre Réal and Joly, Dany and Tastet, Xavier and Aimez, Vincent and Gauvin, Raynald},
	year = {2007},
	pmid = {17455283},
	pages = {92--101},
}

@article{besombes_acoustic_2001,
	title = {Acoustic phonon broadening mechanism in single quantum dot emission},
	volume = {63},
	url = {https://link.aps.org/doi/10.1103/PhysRevB.63.155307},
	doi = {10.1103/PhysRevB.63.155307},
	number = {15},
	urldate = {2023-08-18},
	journal = {Physical Review B},
	author = {Besombes, L. and Kheng, K. and Marsal, L. and Mariette, H.},
	month = mar,
	year = {2001},
	note = {Publisher: American Physical Society},
	pages = {155307},
}

@article{granger_calibration-free_2025,
	title = {Calibration-{Free} {Measurement} of the {Phonon} {Temperature} around a {Single} {Emitter}},
	issn = {1936-0851},
	url = {https://pubs.acs.org/doi/abs/10.1021/acsnano.5c03631},
	doi = {10.1021/acsnano.5c03631},
	urldate = {2025-05-30},
	journal = {ACS Nano},
	author = {Granger, Francis and Bellet-Amalric, Edith and Kheng, Kuntheak and Nogues, Gilles and Ferrand, David and Cibert, Joël},
	month = may,
	year = {2025},
	note = {Publisher: American Chemical Society},
}

@article{brouri_photon_2000,
	title = {Photon antibunching in the fluorescence of individual color centers in diamond},
	volume = {25},
	copyright = {© 2000 Optical Society of America},
	issn = {1539-4794},
	url = {https://opg.optica.org/ol/abstract.cfm?uri=ol-25-17-1294},
	doi = {10.1364/OL.25.001294},
	number = {17},
	urldate = {2025-05-08},
	journal = {Optics Letters},
	author = {Brouri, Rosa and Beveratos, Alexios and Poizat, Jean-Philippe and Grangier, Philippe},
	month = sep,
	year = {2000},
	note = {Publisher: Optica Publishing Group},
	pages = {1294--1296},
}

@article{gosain_quantitative_2022,
	title = {Quantitative analysis of the blue-green single-photon emission from a quantum dot in a thick tapered nanowire},
	volume = {106},
	url = {https://link.aps.org/doi/10.1103/PhysRevB.106.235301},
	doi = {10.1103/PhysRevB.106.235301},
	number = {23},
	urldate = {2024-03-15},
	journal = {Physical Review B},
	author = {Gosain, Saransh Raj and Bellet-Amalric, Edith and Robin, Eric and Den Hertog, Martien and Nogues, Gilles and Cibert, Joël and Kheng, Kuntheak and Ferrand, David},
	month = dec,
	year = {2022},
	note = {Publisher: American Physical Society},
	pages = {235301},
}

@article{geraci_family_2020,
	title = {A family of linear mixed-effects models using the generalized {Laplace} distribution},
	volume = {29},
	issn = {0962-2802},
	url = {https://doi.org/10.1177/0962280220903763},
	doi = {10.1177/0962280220903763},
	number = {9},
	urldate = {2025-05-08},
	journal = {Statistical Methods in Medical Research},
	author = {Geraci, Marco and Farcomeni, Alessio},
	month = sep,
	year = {2020},
	note = {Publisher: SAGE Publications Ltd STM},
	pages = {2665--2682},
}

@article{moreau_quantum_2001,
	title = {Quantum {Cascade} of {Photons} in {Semiconductor} {Quantum} {Dots}},
	volume = {87},
	url = {https://link.aps.org/doi/10.1103/PhysRevLett.87.183601},
	doi = {10.1103/PhysRevLett.87.183601},
	number = {18},
	urldate = {2025-05-08},
	journal = {Physical Review Letters},
	author = {Moreau, E. and Robert, I. and Manin, L. and Thierry-Mieg, V. and Gérard, J. M. and Abram, I.},
	month = oct,
	year = {2001},
	note = {Publisher: American Physical Society},
	pages = {183601},
}

@article{tizei_spatially_2013,
	title = {Spatially {Resolved} {Quantum} {Nano}-{Optics} of {Single} {Photons} {Using} an {Electron} {Microscope}},
	volume = {110},
	url = {https://link.aps.org/doi/10.1103/PhysRevLett.110.153604},
	doi = {10.1103/PhysRevLett.110.153604},
	number = {15},
	urldate = {2025-05-09},
	journal = {Physical Review Letters},
	author = {Tizei, L. H. G. and Kociak, M.},
	month = apr,
	year = {2013},
	note = {Publisher: American Physical Society},
	pages = {153604},
}

@article{bourrellier_bright_2016,
	title = {Bright {UV} {Single} {Photon} {Emission} at {Point} {Defects} in h-{BN}},
	volume = {16},
	issn = {1530-6984},
	url = {https://doi.org/10.1021/acs.nanolett.6b01368},
	doi = {10.1021/acs.nanolett.6b01368},
	number = {7},
	urldate = {2025-05-09},
	journal = {Nano Letters},
	author = {Bourrellier, Romain and Meuret, Sophie and Tararan, Anna and Stéphan, Odile and Kociak, Mathieu and Tizei, Luiz H. G. and Zobelli, Alberto},
	month = jul,
	year = {2016},
	note = {Publisher: American Chemical Society},
	pages = {4317--4321},
}

@phdthesis{gosain_room_2021,
	type = {Ph.{D}. thesis},
	title = {Room temperature single-photon source based on semiconductor quantum-dot nanowire for integrated photonics},
	url = {https://theses.hal.science/tel-03551997},
	urldate = {2023-02-16},
	school = {Université Grenoble Alpes [2020-....]},
	author = {Gosain, Saransh Raj},
	month = dec,
	year = {2021},
}

@article{kikkawa_optical_2022,
	title = {Optical and acoustic phonon temperature measurements using electron nanoprobe and electron energy loss spectroscopy},
	volume = {106},
	url = {https://link.aps.org/doi/10.1103/PhysRevB.106.195431},
	doi = {10.1103/PhysRevB.106.195431},
	number = {19},
	urldate = {2025-05-09},
	journal = {Physical Review B},
	author = {Kikkawa, Jun and Kimoto, Koji},
	month = nov,
	year = {2022},
	note = {Publisher: American Physical Society},
	pages = {195431},
}

@article{gosain,
	title = {The onset of tapering in the early stage of growth of a nanowire},
	volume = {33},
	issn = {0957-4484},
	url = {https://dx.doi.org/10.1088/1361-6528/ac5cfa},
	doi = {10.1088/1361-6528/ac5cfa},
	language = {},
	number = {25},
	urldate = {2023-02-16},
	journal = {Nanotechnology},
	author = {Gosain, Saransh Raj and Bellet-Amalric, Edith and Hertog, Martien den and André, Régis and Cibert, Joël},
	month = apr,
	year = {2022},
	pages = {255601},
}

@article{LI2019220,
title = {Proof-of-principle demonstration of quantum key distribution with seawater channel: towards space-to-underwater quantum communication},
journal = {Optics Communications},
volume = {452},
pages = {220-226},
year = {2019},
issn = {0030-4018},
doi = {https://doi.org/10.1016/j.optcom.2019.07.037},
url = {https://www.sciencedirect.com/science/article/pii/S0030401819306297},
author = {Dong-Dong Li and Qi Shen and Wei Chen and Yang Li and Xuan Han and Kui-Xing Yang and Yu Xu and Jin Lin and Chao-Ze Wang and Hai-Lin Yong and Wei-Yue Liu and Yuan Cao and Juan Yin and Sheng-Kai Liao and Ji-Gang Ren}
}

@article{senellart_high-performance_2017,
	title = {High-performance semiconductor quantum-dot single-photon sources},
	volume = {12},
	copyright = {2017 Nature Publishing Group, a division of Macmillan Publishers Limited. All Rights Reserved.},
	issn = {1748-3395},
	url = {https://www.nature.com/articles/nnano.2017.218},
	doi = {10.1038/nnano.2017.218},
	language = {},
	number = {11},
	urldate = {2023-03-29},
	journal = {Nature Nanotechnology},
	author = {Senellart, Pascale and Solomon, Glenn and White, Andrew},
	month = nov,
	year = {2017},
	pages = {1026--1039},
}

@article{arakawa2020,
	title = {Progress in quantum-dot single photon sources for quantum information technologies: {A} broad spectrum overview},
	volume = {7},
	shorttitle = {Progress in quantum-dot single photon sources for quantum information technologies},
	url = {https://aip.scitation.org/doi/10.1063/5.0010193},
	doi = {10.1063/5.0010193},
	number = {2},
	urldate = {2023-02-16},
	journal = {Applied Physics Reviews},
	author = {Arakawa, Yasuhiko and Holmes, Mark J.},
	month = jun,
	year = {2020},
	pages = {021309},
}

@article{laferriere_position-controlled_2023-1,
	title = {Position-{Controlled} {Telecom} {Single} {Photon} {Emitters} {Operating} at {Elevated} {Temperatures}},
	volume = {23},
	issn = {1530-6984},
	url = {https://doi.org/10.1021/acs.nanolett.2c04375},
	doi = {10.1021/acs.nanolett.2c04375},
	number = {3},
	urldate = {2023-03-28},
	journal = {Nano Letters},
	author = {Laferriére, Patrick and Haffouz, Sofiane and Northeast, David B. and Poole, Philip J. and Williams, Robin L. and Dalacu, Dan},
	month = feb,
	year = {2023},
	pages = {962--968},
}

@article{Donatini2010,
doi = {10.1088/0957-4484/21/37/375303},
url = {https://doi.org/10.1088/0957-4484/21/37/375303},
year = {2010},
month = {aug},
publisher = {},
volume = {21},
number = {37},
pages = {375303},
author = {Donatini, Fabrice and Dang, Le Si},
title = {A single-step electron beam lithography of buried nanostructures using cathodoluminescence
imaging and low temperature},
journal = {Nanotechnology}
}

@article{donatini2018exciton,
  title={Exciton diffusion coefficient measurement in ZnO nanowires under electron beam irradiation},
  author={Donatini, Fabrice and Pernot, Julien},
  journal={Nanotechnology},
  volume={29},
  number={10},
  pages={105703},
  year={2018},
  publisher={IOP Publishing}
}

\end{document}